\documentclass[prl,aps, preprint]{revtex4}
\usepackage{revsymb,amssymb,amscd}
\usepackage{graphicx,natbib}
\usepackage{bm}

\def\be{\begin{equation}}
\def\ee{\end{equation}}
\def\bea{\begin{eqnarray}}
\def\eea{\end{eqnarray}}

\def\>{\rangle}
\def\<{\langle}

\def\tr{\mbox{\rm tr}\,}

\def\lb{\left[}
\def\rb{\right]}

\def\lbL{\lb\rule{0pt}{2.4ex}}

\def\<{\langle}
\def\>{\rangle}

\def\bfig{\begin{figure}[htbp]}
\def\efig{\end{figure}}
\def\bcen{\begin{center}}
\def\ecen{\end{center}}

\def\kua{|\uparrow\>}
\def\kda{|\downarrow\>}

\begin{document}

\title{NMR Quantum Information Processing}

\author{Chandrasekhar Ramanathan, Nicolas Boulant, Zhiying Chen and David G. Cory}
\affiliation{Department of Nuclear Engineering, Massachusetts Institute of Technology, Cambridge, MA 02139} 
\author{Isaac Chuang and Matthias Steffen}
\affiliation{\mbox{MIT Center for Bits and Atoms \& Department of Physics}, Massachusetts Institute of Technology, Cambridge, MA 02139 }
 
\begin{abstract}
Nuclear Magnetic Resonance (NMR) has provided a valuable experimental testbed for quantum information processing (QIP).  Here, we briefly review the use of nuclear spins as qubits, and discuss the current status of NMR-QIP.   Advances in the techniques available for control are described along with the various implementations of quantum algorithms and quantum simulations that have been performed using NMR.   The recent application of NMR control techniques  to other quantum computing systems are reviewed before concluding with a description of the efforts currently underway to transition to solid state NMR systems that hold promise for scalable architectures.
\end{abstract}

\maketitle
\section{Introduction}
Nuclear spins feature prominently in most condensed matter proposals for quantum 
computing \cite{c1,c2,c3,c4,c5,c6,c7,c8,c9,c10}, either directly being used as computational or storage qubits, or being 
important sources of decoherence.  Fortunately, the coherent control of nuclear spins has 
a long and successful history driven in large part by the development of nuclear magnetic 
resonance (NMR) techniques in biology, chemistry, physics, and medicine \cite{c11,c12}.  The 
central feature of NMR that makes it amenable to quantum information processing (QIP) 
experiments is that, in general, the spin degrees of freedom are separable from the other 
degrees of freedom in the systems studied, both in the liquid and solid state.  We can 
therefore describe the Hamiltonian of the spin system quite accurately; there is an 
extensive literature on methods to control nuclear spins, and the hardware to implement 
such control is quite precise.

This readily accessible control of nuclear spins has led to liquid state NMR being used as 
a testbed for QIP, as well as to preliminary studies of potentially scalable approaches to 
QIP based on extensions of solid state NMR.  The liquid state NMR QIP testbed, 
although it is not scalable, has permitted studies of control and QIP in Hilbert spaces 
larger than are presently available with other modalities, and has helped to provide 
concrete examples of QIP.  Here we review what has been learnt in these initial studies 
and how they can be extended to the solid state where scalable implementations of QIP appear to be possible.

The DiVincenzo criteria \cite{DiVincenzo00a} for quantum computation
provide a natural starting place to understand why NMR is such a good
testbed for QIP, and in particular, for implementing quantum
algorithms using liquid state NMR techniques.  These criteria concern
(1) the qubits, (2) the initial state preparation, (3) the coherence
times, (4) the logic gates, and (5) the readout mechanism.

\subsection{Quantum Bits}

Protons and neutrons are elementary particles which carry spin-$1/2$,
meaning that in a magnetic field $\vec{B}$, they have energy
$-\vec{\mu}\cdot \vec{B}$, where the magnetic moment $\vec{\mu} = \mu_N
\vec{I}$ ($\vec{I}$ is the spin operator) is quantized into two energy states, $|\downarrow\>$ and
$|\uparrow\>$ .  These states have an energy
scale determined by the nuclear Bohr magneton $\mu_N = eh/2m_N \approx
5.1 \times 10^{-27}$ A$/$m$^2$ (Table~\ref{tab:nmr-spins}).  Since  spin  is
inherently a {\em discrete} quantum property which exists inside a
finite Hilbert space, spin-$1/2$ systems are excellent quantum
bits.

\begin{table}[htbp]
\begin{center}
\begin{tabular}{|c|c|c|c|c|c|}
\hline\hline
$^1$H     &
$^{19}$F  &
$^{31}$P  &
$^{13}$C  &
$^{29}$Si &
$^{15}$N  
\\
$500$ MHz
& $470$ MHz
& $202$ MHz
& $125$ MHz
& $99$ MHz 
& $50$ MHz 
\\
\hline\hline
\end{tabular}
\end{center}
\caption{Atoms with spin-1/2 nuclei typically used in NMR, and their
energy scales, expressed as a resonance frequency.  Frequencies are
given for $|\vec{B}|\sim 11.74$ Tesla.}
\label{tab:nmr-spins}
\end{table}

Nuclear spins used in NMR QIP are typically the spin-1/2 nuclei of $^1$H,
$^{13}$C, $^{19}$F, $^{15}$N, $^{31}$P, or $^{29}$Si atoms, but
higher order spins such as spin-$3/2$ and
$5/2$ have also been experimentally investigated.  In liquid-state
NMR, these atoms are parts of molecules dissolved in a solvent, such
that the system is typically extremely well approximated as being
$O(10^{18})$ independent molecules.  
Each molecule is an $N$-spin system, with $N$ magnetically distinct nuclei. 
Typically, this molecule sits in a strong, static magnetic field, $B_0$, oriented
along the $\hat z$ axis, such that the $N$ spins precess about $\hat
z$.  The spins interact with each other indirectly via inter-atomic electrons
sharing a Fermi contact interaction with two (or more) nuclei.  The
connectivity of the chemical bonds thus determines which nuclei
interact.

Since the energy scale of the interactions is weak compared with
typical values of $\mu_N |\vec{B}|$, qubits in molecules can be
independently manipulated, and provide a natural tensor product
Hilbert space structure.  This structure is essential for quantum
computing, and in particular, system scalability.

\subsection{Initial state preparation}

The energy scale of a nuclear spin in typical magnetic fields is much
smaller than that of room temperature fluctuations.  As seen in Table~\ref{tab:nmr-spins}, at
$11.74$ T the proton has an NMR resonance frequency of $500$
MHz, whereas room temperature thermal fluctuations are $k_B T \approx
25$ meV $\approx 6$ THz, about $10^4$ times larger.  
As the Boltzmann distribution governs the thermal equilibrium
state of the spins
\be
	\rho = \frac{\exp\lbL{ -\frac{\mu_N |\vec{B}|}{k_B T}}\rb}
			{\cal Z}
\,,
\ee
where ${\cal Z}$ is the partition function normalization factor, $\rho \approx $ {\bf 1} for
$k_B T \gg \mu_N |\vec{B}|$.  Thus, the room  temperature thermal
equilibrium state is a very highly random distribution, with spins
being in their $\kda$ and $\kua$ states with nearly equal probability.

Such a highly mixed state is not ordinarily suitable for quantum
computation, which ideally works with a system initialized to a fiducial state such as
$|00\cdots 0\>$.  It was the discovery of a set of procedures to
circumvent this limitation, which made NMR quantum information processing
feasible and interesting \cite{Cory97a,Gershenfeld97a}.  The essential observation 
is that a $computational$ procedure can be
applied to $\rho$, such that the only observed signal comes from the
net excess population in the $|00\cdots 0\>$ state of the thermal
ensemble.  One class of such techniques averages away the signal
  from all other states.  This averaging can be performed 
  sequentially in time using sequences of pulses which symmetrically
  permute undesired states, spatially using
magnetic field  gradients which prepare spins differently in different parts
  of a single sample, or by selecting a special subset
  of spins depending on the logical state ($|0\>$ or $|1\>$) of
  the unselected spins.  These techniques, known as temporal, spatial, and logical
labeling, do not scale well, and only create a signal strength which
decreases exponentially with the number of qubits realized \cite{Warren97}.

Another class of techniques is based on efficient
compression \cite{Schulman98a}, and in contrast, the signal strength
obtained is constant with increasing number of qubits realized.
Indeed, only $O({\rm poly}(n))$ space and time resources are needed to
initialize $O(n)$ spins using this method, which has now been
experimentally demonstrated \cite{Chang01a}, but there is a constant
overhead factor which prevents it from being practical until $n$ is
large, or the initial temperature of the spins can be brought lower by
several orders of magnitude.  

\subsection{Coherence times}

Nuclear spins couple very weakly with the external world, primarily
due to their small magnetic moment, and the weakness of long-range
magnetic forces.  Thus, typical nuclei in liquid-state molecules may
have a $T_1$ timescale for energy relaxation of between $1$ and $30$
seconds, and a $T_2$ timescale for phase randomization of between
$0.1$ and $10$ seconds. Decoherence may occur due to the presence of quadrupolar nuclei
such as $^{35}$Cl and $^2$D, chemical shift anisotropies, fluctuating
dipolar interactions, and other higher order effects.  Though the coherence lifetimes
are long, the number of gates that can be implemented is limited by the relatively weak
strength of the qubit-qubit couplings (typically a few hundred Hz at most).

\subsection{Logic gates}

In order to perform arbitrary quantum computations only a finite set
of logic gates is required, similar to arbitrary classical
computations. One such set consists of arbitrary single qubit
rotations and the two-qubit controlled-{\sc not} gate. We describe how
each of these is implemented.

The Hamiltonian describing a 2-spin system in an external field $B_0\hat{z}$ 
is (setting $\hbar=1$)
\begin{equation} {\cal H} =  \omega_AI_{zA} + 
\omega_BI_{zB} + {\cal H}_{A,B} \; .
\end{equation} 
 Here, $I_z$ is the spin angular momentum operator in the $\hat z$ direction,
and $\omega_i = -\gamma_i (1-\alpha_i) B_0$, where $\gamma_i$ is the gyromagnetic
ratio for spin $i$, which depends on the nuclear species, $\alpha_i$ is the shielding constant, and ${\cal H}_{A,B}$ is the spin-spin coupling.   The shielding constant depends on the local chemical environment of the nuclei, which shields the local magnetic field, resulting in a shift in frequency by an amount known as the chemical shift such that spins of the same type (e.g.\ protons) can have different resonance
frequencies. That spins have different resonance
frequencies is an important requirement because it permits frequency-dependent addressing of
single qubits.

Spins are manipulated by applying a much smaller radio-frequency (RF)
field, $B_1$, in the $\hat x$-$\hat y$ plane to excite the spins at
their resonance frequencies $\omega_i$. In the rotating frame, to good
approximation, the spin evolves under an effective field $\vec{B}=B_1
\cos(\phi)\hat{x} + B_1 \sin(\phi)\hat{y}$ (where $\phi$ is the RF phase).   
The rotation angle and axis (in the transverse plane) can be controlled by varying 
$\phi$, the magnitude of $B_1$ and the duration of the RF.    Since it is possible to  
generate arbitrary rotations about the $\hat z$-axis using combinations of rotations 
about the $\hat x$ and $\hat y$-axis, it is possible to implement arbitrary single-qubit rotations
using RF pulses. 

\begin{figure}
\centerline{\includegraphics[scale=0.75]{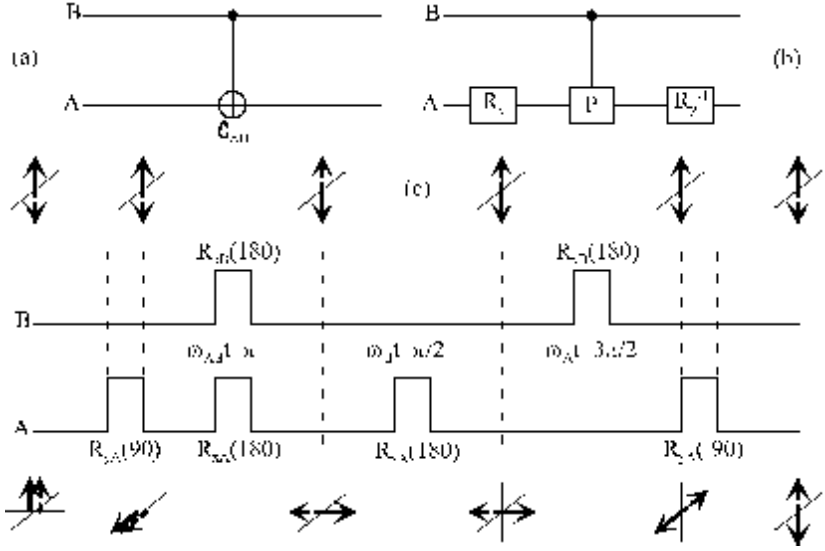}}
\vspace*{2ex}
\caption{(a) A controlled--{\sc NOT} gate acting on two qubits, (b) the
 controlled--{\sc NOT} gates implemented by a controlled phase shift gate
 (specified by a unitary matrix with diagonal elements $\{1,1,1,-1\}$)
 preceded and followed by $\pi/2$ rotations, and (c) the pulse sequence
 and spin orientations corresponding to the components in (b).  Note
 that, unlike a conventional NMR selective population transfer
 sequence, extra refocusing is required to preserve the Bell basis
 exchange symmetry between $A$ and $B$.  The $\hat z$ rotations are implemented via $\hat x$ and $\hat  y$-rotations, which are not explicitly shown.}
\label{fig:cnpulses}
\end{figure}

Two-qubit gates, such as the controlled-{\sc not} gate require
spin-spin interactions. These occur through two dominant mechanisms;
direct dipolar coupling, and indirect through-bond
interactions. The dipolar coupling between two spins is described by an interaction
Hamiltonian of the form
\begin{equation}
{\cal H}_{A,B}^D = \frac{\gamma_A \gamma_B}{r^3} \left( \vec{I}_A \cdot \vec{I}_B - 3(\vec{I}_A \cdot \hat{n})(\vec{I}_B \cdot \hat{n}) \right),
\end{equation}
where $ \hat{n}$ is the unit vector in the direction joining the two
nuclei, and $ \vec{I}$ is the magnetic moment vector. While
dipolar interactions are rapidly averaged away in a liquid, they play a
significant role in liquid crystal \cite{Yannoni99a,Marjanska00a} and solid state NMR QIP experiments. 
Through-bond electronic interactions are an indirect interaction, also known simply
as the scalar-coupling, and take on the form
\begin{equation}
{\cal H}_{A,B}^J =  2\pi J \vec{I}_A \cdot \vec{I}_B =  2\pi J I_{zA} I_{zB} + \pi J \left( I_{A+}I_{B-} + I_{A-}I_{B+} \right),
\end{equation}
where $J$ is the scalar coupling constant. This
interaction is often resolved in liquids.  For heteronuclear species (such that the matrix element of the
$I_{A+}I_{B-} + I_{A-}I_{B+}$ term is small, when $2\pi J \ll \omega_A-\omega_B$), the scalar coupling reduces to 
\begin{equation}
{\cal H}_{A,B}^J \approx  2\pi J I_{zA} I_{zB}.
\end{equation}
Multiple-qubit interactions, such as the controlled-{\sc not}
({\sc cnot}) operation, may be performed by inserting waiting periods
between pulses so that J-coupled evolution can occur. For the
J-coupled two-spin system, a {\sc cnot} can be implemented as a
controlled phase shift preceded and followed by rotations, given by
the sequence ${\cal C}_{AB} =
R_{yA}(270=-90)R_{zB}R_{zA}(-90)R_{zAB}(180)R_{yA}(90)$. This is shown
schematically in Fig.~\ref{fig:cnpulses}.  

In summary, one and two-qubit gates are implemented by applying a sequence of 
 RF pulses interlaced with waiting periods. In this
sense, it is perhaps interesting to note that the sequence of
elementary instructions (pulses and delay times) are the machine
language of the NMR quantum information processor.

\subsection{Readout mechanism}

Readout of the quantum state in NMR QIP is not the usual ideal
projective von Neumann measurement.  Instead, the system is
continually read out by the weak coupling of the magnetic dipole
moments to an external pickup coil, across the ends of which a voltage
is produced by Faraday induction.  This coil is usually the same coil
as that used to produce the strong RF pulses which control the spins,
and thus it only detects magnetization in the $\hat{x}-\hat{y}$ plane.
The induced voltage, known as the free induction decay, may be
expressed as
\be
        V(t) = -2V_0 \tr \lbL{ e^{-i{\cal H} t} \rho e^{i{\cal H} t} 
                        (iI_x^k+I_y^k) }\rb
\,,
\end{equation}
where $\cal{H}$ is the Hamiltonian for the spin system, $I_x^k$ and
$I_y^k$ operate only on the $k^{th}$ spin, and $V_0$ is a constant
factor dependent on coil geometry, quality factor, and maximum
magnetic flux from the sample volume. 

\begin{figure}
\begin{center}
\includegraphics*[width=9cm]{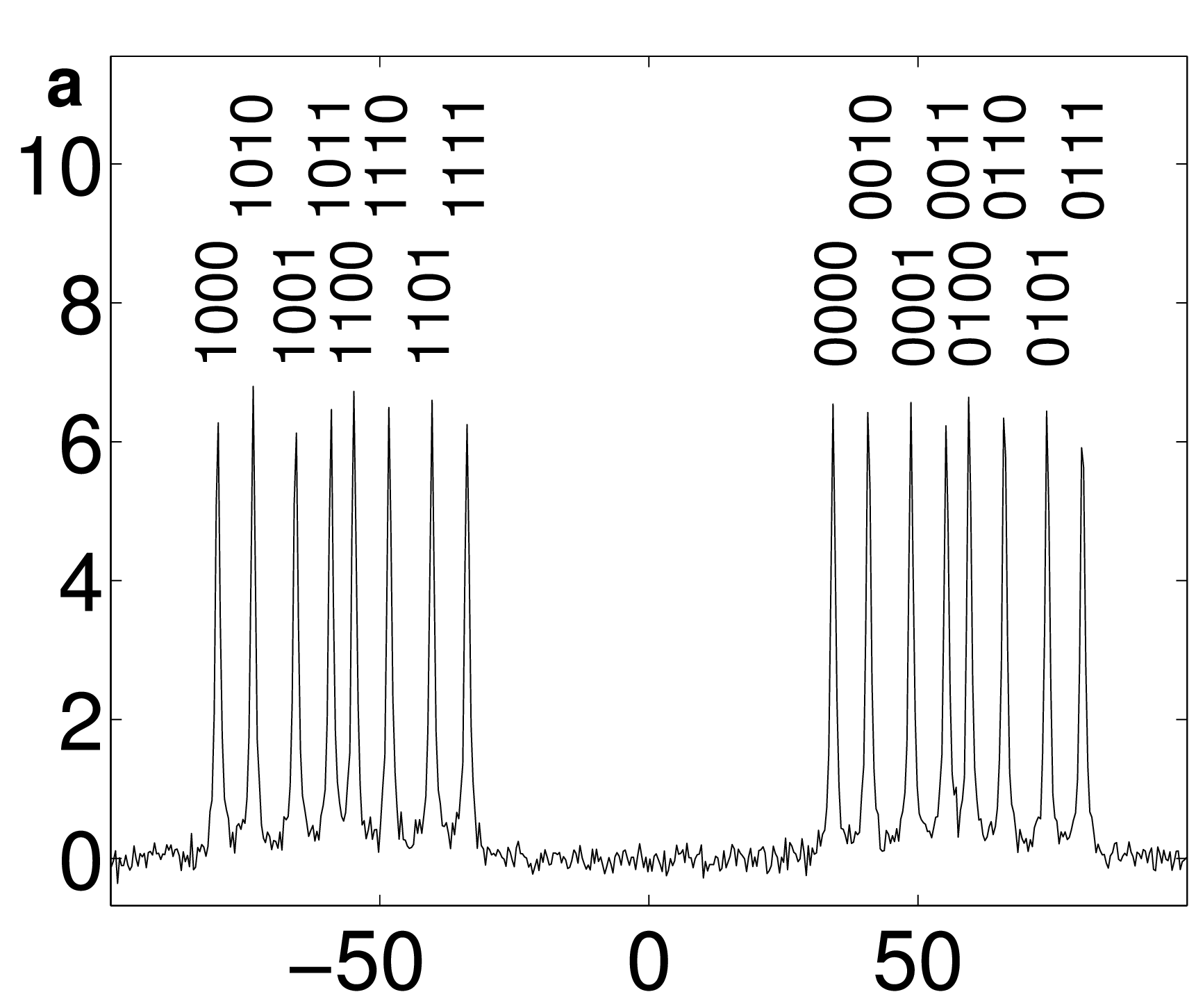}
\end{center}
\caption{Thermal equilibrium spectrum of a $5$ spin
molecule.  Each peak corresponds to a certain logical
  state of the remaining four spins, which is indicated by the binary
  numbers. The real part of the spectrum is shown, in arbitrary
  units. Frequencies are given with respect to $\omega_i/2\pi$, in
  Hertz \protect\cite{Vandersypen00b}.}
\label{fig:nmrspect}
\end{figure}

The Fourier transform of $V(t)$ is the NMR spectrum as
shown in Fig.~\ref{fig:nmrspect} for example.  When properly calibrated, the NMR spectrum immediately reveals the
logical state of qubits which are either $|0\>$ or $|1\>$.
Specifically, for example, if the initial state of a two-spin
$^1$H-$^{13}$C system is described by a diagonal density matrix,
\be
          \rho = \left[ 
			\begin{array}{cccc}
				a & 0 & 0 & 0 \\
                                0 & b & 0 & 0 \\
                                0 & 0 & c & 0 \\
                                0 & 0 & 0 & d 
			\end{array}
			 \right]
\end{equation}
(where the states are $00$, $01$, $10$, and $11$, with proton on left
and carbon on right and $a$, $b$, $c$, and $d$ denote the occupation probabilities) then after a $R_x(\pi/2)$ readout pulse, the
integrals of the two proton peaks (in the proton frequency spectrum)
are given by $a-c$ and $b-d$, and the integrals of the two carbon
peaks are given by $a-b$ and $c-d$.  Both the proton and carbon spectra contain 
two peaks because of the J-coupling interaction during the measurement period.

One important issue in the readout of QIP results from
NMR arises because the system is an {\em ensemble}, rather than a
single $N$-spin molecule.  The problem is that the output of a typical
quantum algorithm is a random number, whose distribution gives
information which allows the problem to be solved.  However, the
average value of the random variable would give no relevant
information, and this would be the output if the quantum algorithm
were executed without modification on an NMR quantum computer.

This problem may be resolved \cite{Gershenfeld97a} by appending an
additional computational step to the quantum algorithm to eliminate or
reduce the randomness of its output.  For example, the output of
Shor's algorithm is a random rational number $c/r$, from which
classical post-processing is usually employed to determine a number
$r$, which is the period of the modular exponentiation function under
examination.  However, the post-processing can just as well be
performed on the quantum computer itself, such that $r$ is determined
on each molecule separately.  From $r$, the desired prime factors can
also be found, and tested; only when successful does a molecule
announce an answer, so that the ensemble average reveals the factors.
Similar modifications can be made to enable proper functioning of all
known exponentially fast quantum algorithms~\cite{Deutsch92,Simon94,Kitaev95} 
on an NMR quantum computer.

\section{Quantum Control}  
Implementing an algorithm on a quantum computer requires performing both unitary 
transformations and measurements.  Errors in the control and the presence of noise can 
severely compromise the accuracy with which a unitary transform can be implemented.  
Quantum control techniques are used to maximize the accuracy with which such 
operations can be performed, given some model for the system's dynamics.  NMR has 
provided valuable insight into the design of schemes to control quantum systems, as the 
task of applying pulse sequences to perform operations that are selective, and also robust 
against experimental imperfections, has been the subject of extensive studies \cite{Jones03b,c14}.
A single, isolated quantum system will evolve unitarily under the Hamiltonian of the 
system.  In an ensemble measurement (whether in space or time), an isolated system can 
also appear to undergo non-unitary dynamics, called incoherent evolution, due to a 
distribution of fields over the ensemble \cite{c15,c16}.  An open quantum system, interacting with an 
environment, will decohere if these interactions are not controlled.  The study of quantum 
control using NMR can therefore be separated into three different subsections: 

\begin{itemize}
\item Coherent control: how can one design RF control schemes to implement the 
correct unitary dynamics for a single, isolated quantum system ?
\item Incoherent noise:  how can coherent control be extended to an ensemble, given 
that the system Hamiltonian will vary across the ensemble ?
\item Decoherent noise: how can one achieve the desired control, when coupling to the 
environment is taken into account ? 
\end{itemize}

\subsubsection{Coherent control}
The density matrix of a closed system evolves according to the Liouville--Von Neumann 
equation of motion:
\begin{equation}
\frac{d\rho}{dt}=-i\left[\cal{H}_{\mathrm{int}} + \cal{H}_{\mathrm{ext}}, \rho \right]
\end{equation}
where ${\cal H}_{\mathrm{int}}$ is the internal Hamiltonian of the system of qubits, and ${\cal H}_{\mathrm{ext}}$ the externally 
applied control fields. More specifically, extending Eqs.\ [2] and [4], the internal Hamiltonian for a system of {$N$ spin-
1/2 nuclei in a large external magnetic field $B_0$ is :
\begin{equation}
{\cal H}_{\mathrm{int}} = \sum_{k=1}^{N} - \gamma_k(1-\alpha_k)B_0(r)I_z^k + 2\pi\sum_{j>k}^{N}\sum_{k=1}^{N} J_{kj}I^k \cdot I^j \; \; 
\end{equation}
where $-\gamma (1-\alpha_k) B_0$(r) is the chemical shift of the $k^{th}$ spin.  The corresponding experimentally-controlled RF Hamiltonian is : 
\begin{equation}
{\cal H}_{\mathrm{ext}} = \sum_{k=1}^{N} -\gamma_k f(r) B_{RF}(t) e^{-i\phi (t)I_z^k}I_x^k e^{i\phi (t)I_z^k} 
\end{equation}
where the time-dependent functions $B_{RF}(t)$ and $\phi(t)$ specify the applied RF control field, 
while $f(r)$ reflects the distribution of RF field strengths over the sample. The spatial 
variation of the static and RF magnetic fields leads to incoherence \cite{c15}.  We will return to 
this in the next section.

If the total Hamiltonian is time-independent, possibly through transformation into a 
suitable interaction frame, the equation of motion can be integrated easily and yields a 
unitary evolution of the density matrix $\rho(t) = U(t)\rho(0)U^{\dag}(t)$. Given an internal 
Hamiltonian and some control resources, how can we implement a given propagator or 
prepare a given state?  In QIP, it is necessary to implement the correct propagator, which 
requires designing gates that perform the desired operation regardless of the input state.  
We will therefore focus our discussion on NMR control techniques that are universal, i.e.\ 
whose performance is essentially independent of the input state, although sequences 
whose performances are state dependent can also be useful for initialization purposes.
Average Hamiltonian theory is a powerful tool for coherent control that was initially 
developed for NMR \cite{c17}.  Waugh and Harberlen have provided a theoretical framework 
to implement a desired effective Hamiltonian evolution of a spin system over some period 
of time.  Such a tool fits well into the context of QIP since it aims to implement the 
correct propagator over the system Hilbert space while refocusing undesired qubit-qubit 
interactions. The basic idea is to apply a cyclic train of pulses $P=\{P_j\}_{j=1}^{M}$ with $\prod P_j=$ {\bf 1} to 
the system which, in its simplest form, are assumed to be infinitely short and equally 
spaced by $\Delta t > 0$.  The net controlled evolution over the period $T=M\Delta t$ can then be 
expressed as 
\begin{equation}
e^{-i\bar{H}T} = \prod_{k=0}^{M}e^{-iH_k \Delta t}
\end{equation}
where the "toggling-frame" Hamiltonians $H_k =U_k^{\dag}H_k U_k$  are expressed in terms of 
the composite pulses $U_k=\prod_{j=1}^k P_j, k=1, \ldots , M, U_0=${\bf 1}.  Any average Hamiltonian (up to a 
scalar multiple) can be implemented in NMR systems of distinguishable spins \cite{c18}.  
This work has been extended to correct for some experimental 
imperfections or uncertainties primarily using symmetry arguments.  Composite pulses 
have also been used to design robust control sequences as they can be designed to be 
self-compensating for small experimental errors \cite{Jones03b,c14,c19,c20,c21,c22,c23,c24}.

Strong modulation of the spin system currently represents the state of the art in 
performing selective, controlled operations in large Hilbert spaces (up to 10 qubits).  
Strong modulation of the spins permits accurate selective rotations while refocusing the 
internal Hamiltonian during the RF irradiation of the spins \cite{c25}.  Figure \ref{fig:fidelities} shows the 
dependence of the fidelities achievable on the available control resources, using this 
technique \cite{c15}.  These are simulation results, assuming only unitary dynamics.  The fidelities
are seen to improve both with improved distinguishability of the spins (higher field strengths), and stronger
modulation (increased RF power). 
The drawback of this technique is that it is not scalable, as the time necessary to find the 
numerical solutions grows exponentially with the number of qubits (see \cite{Vandersypen04a} for a review 
of other techniques).   An alternative, scalable approach that relies on optimization over single and pairs of nuclei has been described \cite{Knill00a}, though it does not appear to work as well.

Finally, there is a growing body of work in the area of optimal control theory for 
quantum systems \cite{c27,c28,c29}, which has also used NMR as an experimental testbed.  It is 
foreseeable that it might be possible to combine the ideas presented in these studies with strong modulation and pulse-shaping techniques to design optimal control 
sequences, given some knowledge about the system decoherence and the control 
parameters.

\begin{figure}
\begin{center}
\includegraphics[scale=0.65]{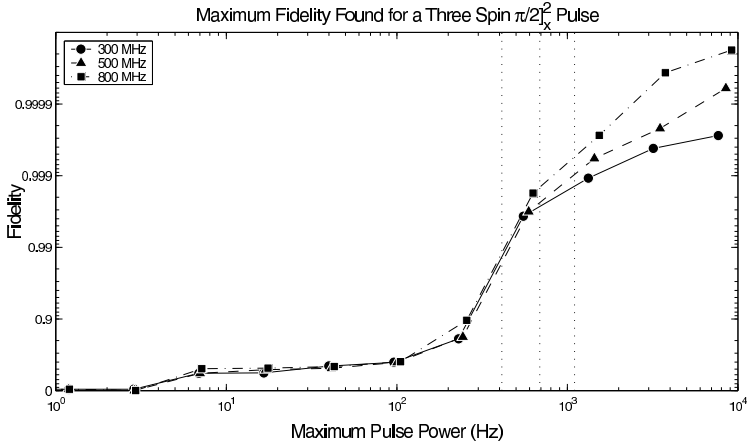}
\caption{Maximum fidelity achievable versus external magnetic field strengths and maximum 
radio-frequency power available for strongly modulating pulses.  The frequencies shown are the proton Larmor frequencies at different magnetic fields (300 MHz $\equiv$ 7 T, 500 MHz $\equiv$ 11.7 T, 800 MHz $\equiv$ 18.8 T).}
\label{fig:fidelities}
\end{center}
\end{figure}

\subsubsection{Incoherent noise}
Incoherent noise arises from a spatial or temporal distribution of experimental parameters 
in an ensemble measurement \cite{c15}.  It is manifested in NMR in the spatial dependence of 
the Hamiltonian ${\cal H}$.  The density matrix at a given location in the sample still obeys the 
Liouville-Von Neumann equation where the internal and external Hamiltonian are now 
spatially dependent.  Since it is the spatially averaged density matrix that is measured, the apparent 
evolution of the ensemble system is non-unitary and yields the following operator sum 
representation \cite{c30} of the superoperator 
\begin{equation}
\rho(t) = \sum_{i} p_iU_i(t)\rho(0)U_i^{\dag}(t)
\end{equation}
where $U_i$ is a unitary operator and $p_i$ represents the fraction of spins that experience a 
given $U_i$ ($\sum_i  p_i = 1$). This incoherent evolution can be counteracted using a different set of 
techniques than those used to deal with the decoherent errors to be discussed in 
the next section.

Hahn's pioneering work showed that inhomogeneities in the Hamiltonian could 
be refocused during an experiment if an external control Hamiltonian orthogonal to the 
inhomogeneous Hamiltonian was available \cite{c32} (see Figure \ref{fig:cp}).  In the case of an 
inhomogeneous but static Hamiltonian, the correlation time of the noise is infinite.  This 
work was extended by Carr and Purcell to counteract long, but not infinite, correlation 
time noise fluctuations due to molecular diffusion \cite{c31}.  Composite pulse sequences have also 
been used to counteract the effects of incoherent processes \cite{c19,c20,c21,c22,c23,c24,c33,c34}, but often 
assume specific input states, or still need to be proven effective over the full Hilbert space 
of multi-qubit systems.  Spin decoupling fits into a similar framework, and provides a 
means of modulating the system in order to average out unwanted interactions with the 
environment \cite{c35}.  It has inspired coherent approaches to other control problems for error 
correction purposes \cite{c36,c37,c38}. 

The use of strongly modulating pulses has been extended to incorporate incoherent 
effects, considering local unitary dynamics over the ensemble \cite{c15}.  {\em A priori} knowledge of the 
inhomogeneity of the external Hamiltonian was used to find robust pulse sequences 
yielding a higher fidelity operation over the ensemble.  This knowledge was easily 
obtained by spectroscopic NMR techniques.  Though this work focused on counteracting 
the main source of incoherent errors in an NMR experiment, i.e.\ RF inhomogeneity, it 
could easily be extended to compensate for experimental uncertainties such as the phase 
noise of the RF irradiation or static $B_0$  field inhomogeneities. 

\begin{figure}
\begin{center}
\includegraphics[scale=0.5]{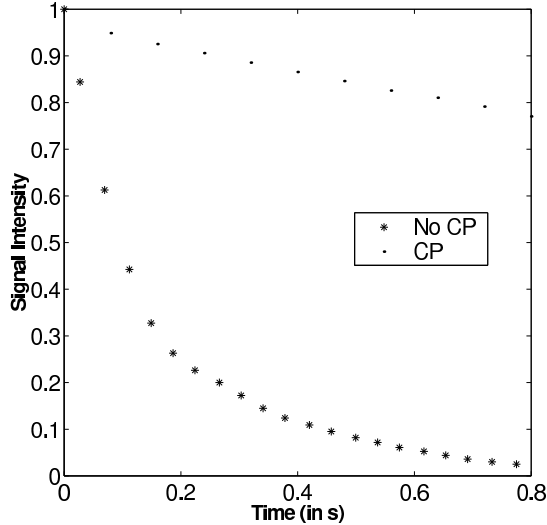}
\caption{Decay of the transverse magnetization of a nuclear spin in a liquid state sample with 
large $B_0$ field inhomogeneities both with and without a Carr-Purcell (CP) sequence.}
\label{fig:cp}
\end{center}
\end{figure}

\subsubsection{Decoherent noise}
If the coupling between the system and the environment is weak enough and the 
correlation time of the noise is short, the evolution of the system is Markovian and obeys 
the following master equation:
\begin{equation}
\frac{d |\rho(r)\rangle}{dt} = -\left(i\hat{H}(r) + \Gamma\right) |\rho(r) \rangle
\end{equation}
where $\Gamma$ is the Liouville space relaxation superoperator \cite{c12}.  This equation yields a 
non-unitary evolution of the density matrix so that pure states can evolve into mixed states.  
To understand and test models of decoherence, methods based on quantum process 
tomography (QPT) were developed to measure $\Gamma$ \cite{c39,c40}, so that the dynamics of the 
system could be simulated more accurately. The full model of the system including 
coherent, incoherent and decoherent dynamics, has been tested extensively with a three 
qubit QPT of the Quantum Fourier Transform superoperator \cite{c41}.  
When knowledge about the noise operators is available, quantum error correction 
(QEC) schemes can in principle be designed to allow quantum computing in the presence 
of imperfect control. NMR has primarily been used to test the ideas of quantum error 
correction (QEC) \cite{c42,c43,c44} and of fault-tolerant quantum computations \cite{c45}.  Experiments 
were carried out to investigate different QEC scenarios \cite{c46,c47,c48,c49,c50,c51}, in addition to encoded 
operations acting on logical qubits \cite{c52}.  Schemes to implement logical encoded quantum 
operations while remaining in a protected subspace have also been investigated \cite{c48} for a 
simple system made of two physical qubits and are still being studied for larger systems.

\section{Quantum Algorithms}

Many quantum algorithms have now been implemented using liquid-state
NMR techniques.  The first quantum algorithms implemented with NMR were Grover's
algorithm \cite{Chuang98a,Jones98b} and the Deutsch-Jozsa algorithm
\cite{Chuang98c,Jones98a} for two qubits.  The quantum counting algorithm 
was implemented soon afterwards using two-qubits \cite{Jones99b}.  A variety of 
implementations of Grover's algorithm and the Deutsch-Jozsa algorithm have subsequently 
been performed.  The two-qubit Grover
search was re-implemented using a subsystem of a three qubit system
\cite{Vandersypen99a}, demonstrating state preparation using logical labeling.  
A three qubit Grover search has been implemented, in which $28$ Grover iterations were
performed, involving $280$ two-qubit gates \cite{Vandersypen00a}.  A
three-qubit Deutsch-Jozsa algorithm using transition selective pulses
\cite{Linden98b}, another more advanced version using {\sc swap} gates
to avoid small couplings \cite{Collins00a}, and yet another
implementation without {\sc swap} gates \cite{Kim00a}.  A subset of a five-qubit 
Deutsch-Jozsa algorithm has also been implemented \cite{Marx00a}.  

The implementation of quantum algorithms reached a new level with the
full implementation of a Shor-type algorithm using five qubits
\cite{Vandersypen00b}.  This work involved the use of exponentiated
permutations, combined with the quantum Fourier transform, which had
been previously been implemented \cite{Weinstein01a}. 

\begin{figure}[htbp]
\begin{center}
\includegraphics*[width=5in]{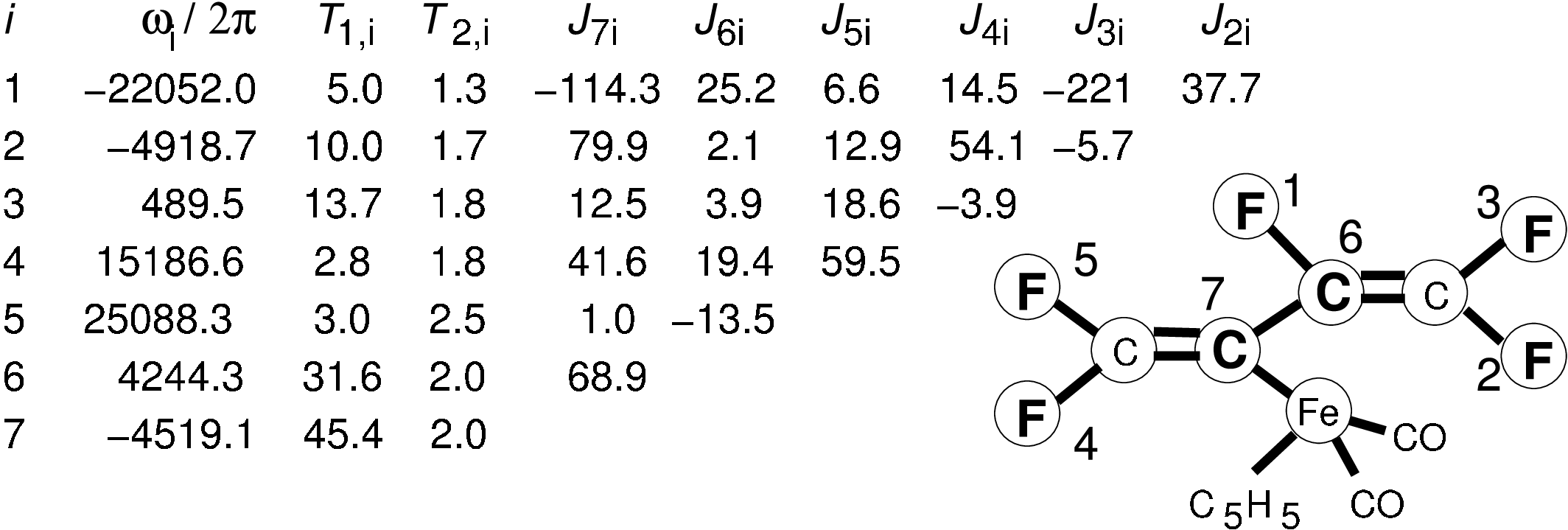}
\end{center}
\vspace*{1ex}
\caption{The seven spin molecule used in the quantum factoring NMR
experiment, showing its J-coupling constants, $T_1$ and $T_2$
relaxation times (in seconds), and chemical shifts (in Hertz) at
$11.74$ Tesla.} 
\label{fig:shormolecule}
\end{figure}

The most complex quantum algorithm realized to-date is the demonstration of 
Shor's algorithm using liquid-state NMR QIP methods:   In
this work \cite{Vandersypen01a}, a seven-qubit molecular system was used
to factor the number $15$ into its prime factors.  This molecule,
shown in Fig.~\ref{fig:shormolecule}, was specially chemically
synthesized to give resolvable fluorine spectra, in which the two
$^{13}$C nuclei, and the five $^{19}$F nuclei, could each be addressed
independently because of the spread of their resonant frequencies.
The NMR spectra of this molecule are quite remarkable; for example,
the thermal spectrum of $^{19}$F spin number 1 shows 64 lines, 
corresponding to the random states of the other 6 spins
(Fig.~\ref{fig:f-spect}).  Several hundred pulses were applied, with a wide variety of phases,
and shapes, at seven different frequencies, in this demonstration of the
factoring algorithm
(Fig.~\ref{fig:shorpulseseq}).  A comparison of the experimental results with numerical simulations
suggests that decoherence was the major source of error in the experiment rather than errors 
in the unitary control, which is remarkable considering the number of pulses applied.

\begin{figure}[htbp]
\begin{center}
\includegraphics*[width=3in]{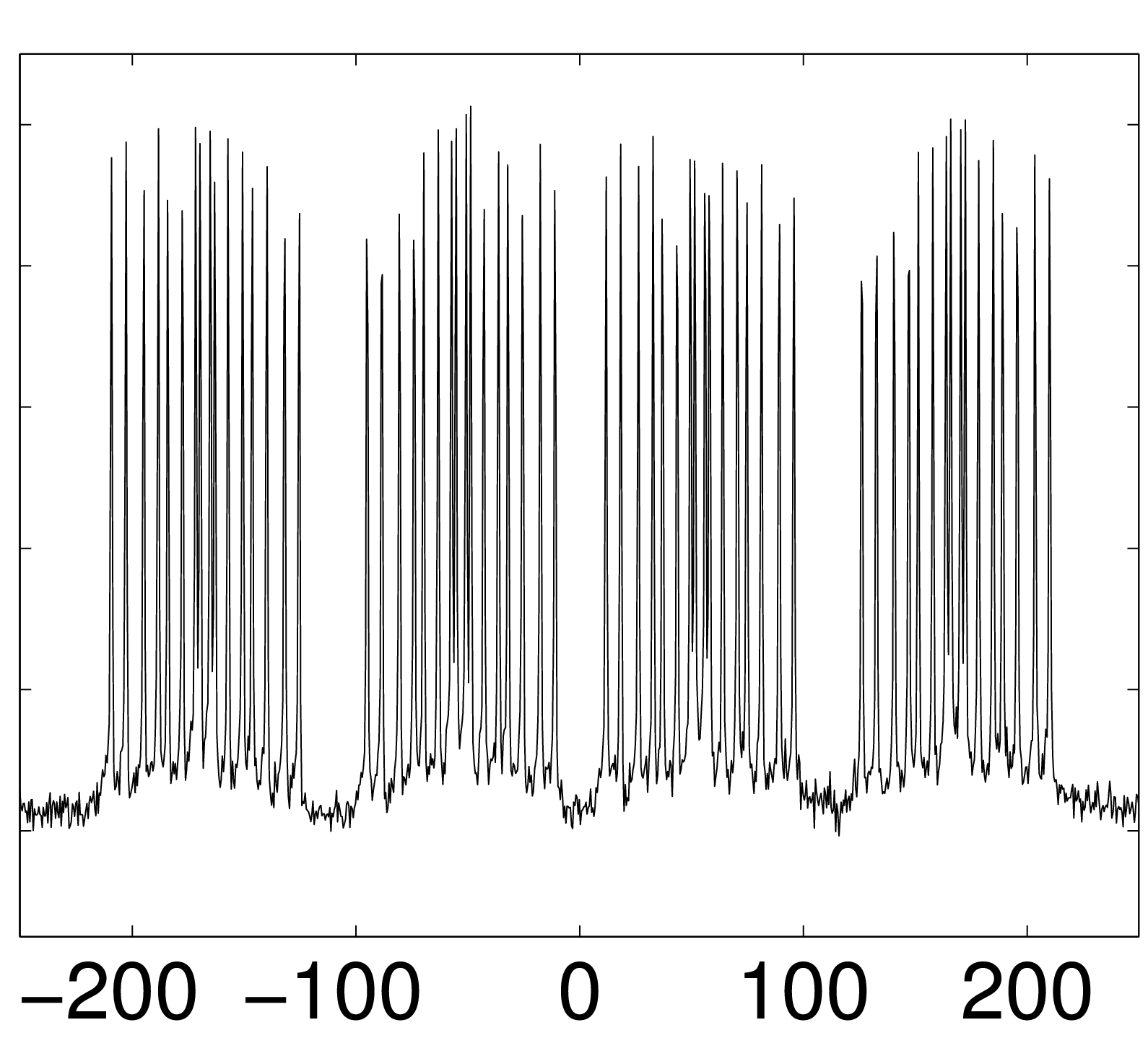}
\end{center}
\caption{Experimentally measured thermal equilibrium spectra of the
NMR spectrum of fluorine atom number 1, in the molecule of
Fig.~\ref{fig:shormolecule}. The real part of the spectrum is shown,
in arbitrary units. Frequencies are given with respect to
$\omega_i/2\pi$, in Hertz.}
\label{fig:f-spect}
\end{figure}

\begin{figure}[htbp]
\begin{center}
\includegraphics*[width=4.8in]{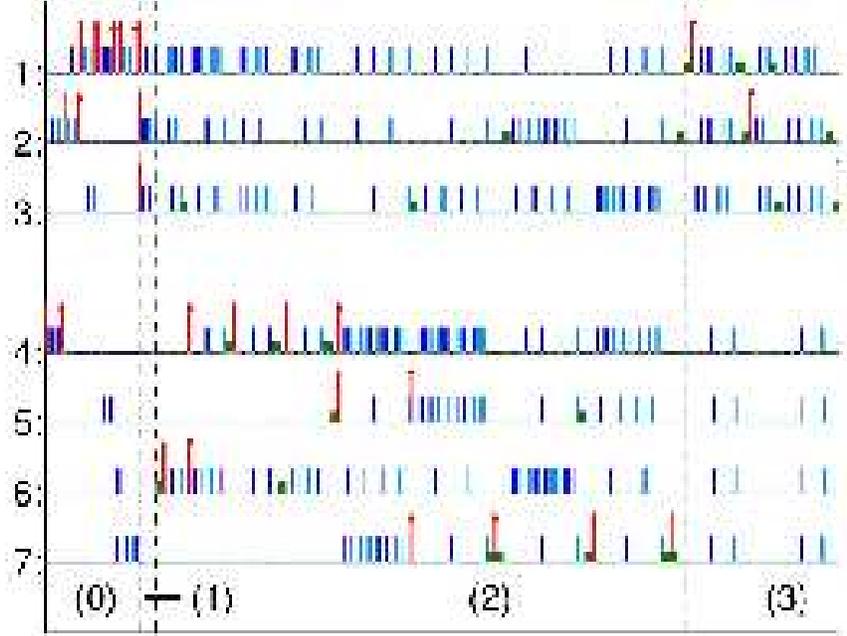}
\end{center}
\vspace*{-4ex}
\caption{Pulse sequence for implementing Shor's algorithm to factor
$N=15$ (for case $a=7$), using seven qubits. The
  four steps $0$ through $3$ correspond to different steps in Shor's
  algorithm.  The tall lines represent
$90^\circ$ pulses selectively acting on one of the seven qubits
(horizontal lines) about positive $\hat{x}$ (no cross), negative
$\hat{x}$ (lower cross) and positive $\hat{y}$ (top cross). Note how
single $90^\circ$ pulses correspond to Hadamard gates and pairs of
such pulses separated by delay times correspond to two-qubit gates.
The smaller lines denote $180^\circ$ selective pulses used for
refocusing, about positive (darker shade) and negative $\hat{x}$
(lighter shade).  Rotations about $\hat{z}$ are denoted by even
smaller and thicker lines and were implemented with frame-rotations.
Time delays are not drawn to scale.  The vertical dashed black lines
visually separate the steps of the algorithm; step (0) shows one of
the 36 temporal averaging sequences.}
\label{fig:shorpulseseq}
\end{figure}

Most recently, among NMR implementations of quantum algorithms, has
been the realization of a three-qubit adiabatic quantum optimization
algorithm \cite{Steffen03a}.  In this work, the $^1$H, $^{13}$C, and
$^{19}$F nuclei in molecules of bromofluoromethane were used as
qubits, and the solution to a combinatorial problem, {\sc maxcut}, was
obtained, using an optimization algorithm proposed by
Farhi \cite{Farhi01a} and Hogg \cite{Hogg00a}.  This algorithm is
notable because it is fundamentally different in nature from Shor-type
quantum algorithms, and may obtain useful speedups for a wide variety
of optimization problems.

\subsubsection{Other quantum protocols}

The Greenberger-Horne-Zeilinger state and its derivative entangled
states of three particles have been studied as well. First, an
effective-pure GHZ state was prepared \cite{Laflamme98a}, and later a
similar experiment was done with seven spins \cite{Knill00a}. The
claim of having created entangled states was later refuted based on
the fact that spins at room temperature are too mixed to be entangled
\cite{Braunstein99a}.  GHZ correlations have since
been further studied on mixed states \cite{Nelson00a}.

 A quantum teleportation protocol was implemented using three qubits \cite{Nielsen98b}. 
Superdense coding has been realized \cite{Fang00a}, and an approximate 
quantum cloning experiment has been implemented (an unknown quantum state cannot be perfectly
copied; it can only be approximately cloned) \cite{Cummins02a}.  The quantum 
Baker's map has also been implemented \cite{Weinstein02a}.

 Several experiments have been performed in an attempt to increase the thermal polarization of nuclear
 spins in liquid solution, as this poses a significant challenge for scaling NMR quantum
computers to many qubits.
An algorithm approach
implementing the basic building block of the Schulman-Vazirani cooling scheme has been
demonstrated \cite{Chang01a}.  High initial polarization of the proton and carbon in a
chloroform molecule have been obtained by transfer from optically pumped
rubidium, through hyperpolarized xenon, and a two-qubit Grover search 
implemented on this non-thermally polarized system
\cite{Verhulst01a}. In a different approach \cite{Hubler00a} {\em
para} hydrogen was transformed into a suitable molecule leading to a
polarization of $10 \%$ which is much larger than the thermal
polarization of $O(10^{-4})$. A quantum algorithm was subsequently
performed on this molecule.  Most recently,  a two-spin system was initialized 
  to an effective purity of $0.916$ by chemically synthesizing a
  two-spin molecule using highly polarized {\em para}-hydrogen \cite{Jones03a}.

\section{Quantum Simulation}
In 1982, Feynman recognized that a quantum system could efficiently be simulated by a 
computer based on the principle of quantum mechanics rather than classical mechanics 
\cite{c53}.  This is perhaps one of the most important short term applications of QIP. An 
efficient quantum simulator will also enable new approaches to the study of multibody 
dynamics and provide a testbed for understanding decoherence.  

A general scheme of simulating one system by another is expressed in Figure \ref{fig:modelsim}.  The 
goal is to simulate the evolution of a quantum system $S$ using a physical system $P$.  The 
physical system is related to the simulated system via an invertible map $\phi$ , which creates 
the correspondence of states and propagators between the two systems. In particular, the 
propagator $U$ in the system $S$  is mapped to $V=\phi U\phi^{-1}$ . After the evolution of the 
physical system from state $p$ to $p_T$, the inverse map brings it back to the final state $s(T)$ of 
the simulated system.

\begin{figure}
\includegraphics[scale=0.5]{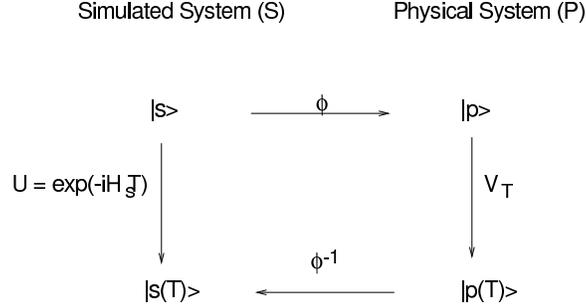}
\caption{Correspondence between the simulated and physical system. The initial state $s$ evolves 
to $s(T)$ under the propagator $U$.  This process is related to the evolution of state $p$ in the physical 
system by an invertible map $\phi$.}
\label{fig:modelsim}
\end{figure}
 
The first explicit experimental NMR realization of such a scheme was the simulation 
of a truncated quantum harmonic oscillator (QHO) \cite{c54}.  The states of the truncated QHO 
were mapped onto a 2-qubit system as follows
\begin{eqnarray}
|n=0 \rangle & \leftrightarrow & |0 \rangle |0 \rangle \equiv |00\rangle \nonumber \\
|n=1 \rangle & \leftrightarrow & |0 \rangle |1 \rangle \equiv |01\rangle \nonumber \\
|n=2 \rangle & \leftrightarrow & |1 \rangle |0 \rangle \equiv |10\rangle \nonumber \\
|n=3 \rangle & \leftrightarrow & |1 \rangle |1 \rangle \equiv |11\rangle \;.
\end{eqnarray}
The propagator of the truncated QHO 
\begin{equation}
U = \exp\left\{-i\left(\frac{1}{2}|0\rangle\langle0| + \frac{3}{2}|1\rangle\langle1|+\frac{5}{2}|2\rangle\langle2| + \frac{7}{2}|3\rangle\langle3|\right)\Omega T\right\}
\end{equation}
($\Omega$ is the oscillator frequency) was mapped onto the following propagator of a two-spin 
system
\begin{equation}
V_T = \exp\left\{i\left(2I_z^2\left(1+I_z^1\right)-2\right)\Omega T\right\} \; .
\end{equation}
 Implementing this propagator on 
the 2-spin system simulates the truncated QHO as shown in Figure \ref{fig:qho}.  
 
\begin{figure}

\vspace*{-0.8in} \hspace*{-0.75in}
\includegraphics[scale=1]{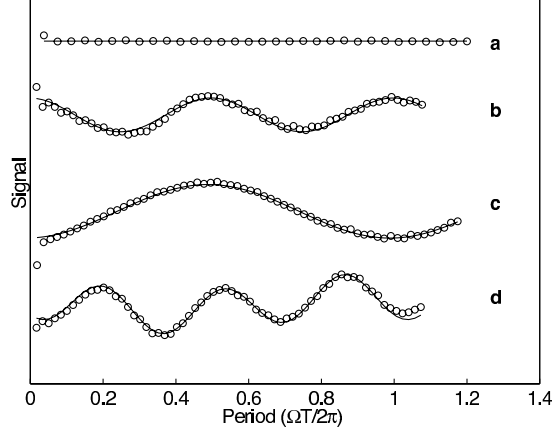}
\vspace*{0.7in}
\caption{NMR signals demonstrate a quantum simulation of truncated harmonic oscillator.  The 
solid lines are fits to theoretical expectations.  Evolution of the different initial states  are shown: 
(a) evolution of $|0\rangle$ with no oscillation (b) evolution of $|0\rangle + i|2\rangle$ , showing $2\Omega$ oscillations (c) evolution of  $|0\rangle + |1\rangle + |2\rangle + |3\rangle$, showing   $\Omega$ oscillation and (d) $3\Omega$  oscillations.}
\label{fig:qho}

\end{figure}

Quantum simulation however is not restricted to unitary dyamics.  It is sometimes 
possible to engineer the noise in a system to control the decoherence behavior and 
simulate non-unitary dynamics of the system \cite{c55}.  Simple models of decoherence have 
been shown using a controlled quantum environment in order to gain further 
understanding in decoherence mechanisms.  In one model \cite{c56}, the environment is taken 
to be a large number of spins coupled to a single system spin so that the total Hamiltonian 
can be expressed as
\begin{equation}
{\cal H} = \omega_1I_z^1 + \sum_{k=2}^{N} \omega_k I_z^k + 2\pi \sum_{k=2}^{N} J_{1k}I_z^1I_z^k
\end{equation}
corresponding to the system, the 
environment, and the coupling between the system and the environment respectively (the 
couplings within the environment were omitted here for simplicity).  Note that the form is 
identical to the weak coupling Hamiltonian of a liquid state NMR sample presented in the 
previous sections.  However, the number of spins in a typical QIP NMR molecule is small, 
which makes the decoherence arising from the few ``system-environment" couplings 
rather ineffective, as the recurrence time due to a small environment is relatively short.  
This can be circumvented by using a second ``classical" environment which interacts with 
a small size quantum environment (see Fig. \ref{fig:dec_model} for an illustration of the model) \cite{c57}. 
\begin{figure}
\includegraphics[scale=0.5]{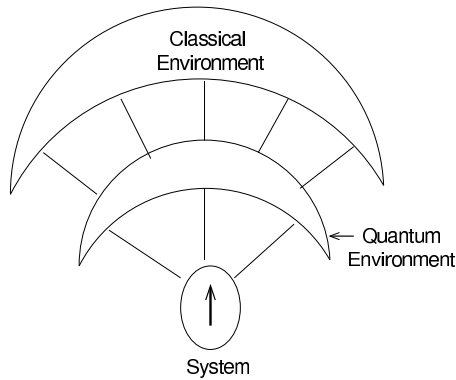}
\caption{Basic model for the system, local quantum and classical environments.}
\label{fig:dec_model}
\end{figure}

In this model, following the evolution of the system and the small quantum 
environment, a random phase kick was applied to the quantum environment.  This has the 
effect of scrambling the system phase information stored in the environment during the 
coupling interaction and therefore emulates the loss of memory.  When the kick angles 
are averaged over small angles, the decay induced by the kicks is exponential and the rate 
is linear in the number of the kicks \cite{c57}.  As the kick angles are completely randomized 
over the interval from $0$ to $2\pi$, a Zeno type effect is observed.  Figure \ref{fig:kicks} shows the 
dependence of the decay rate on the kick frequency: the decay rate initially increases to 
reach a maximum and then decreases, thereby illustrating the motional narrowing \cite{c11} or 
decoupling \cite{c35} limit.  This NMR-inspired model thus provides an implementation of 
controlled decoherence yielding both non-exponential and exponential decays (with some 
control over the decay rates), and can be extended to investigate other noise processes.   

\begin{figure}
\begin{center}
\includegraphics[scale=0.45]{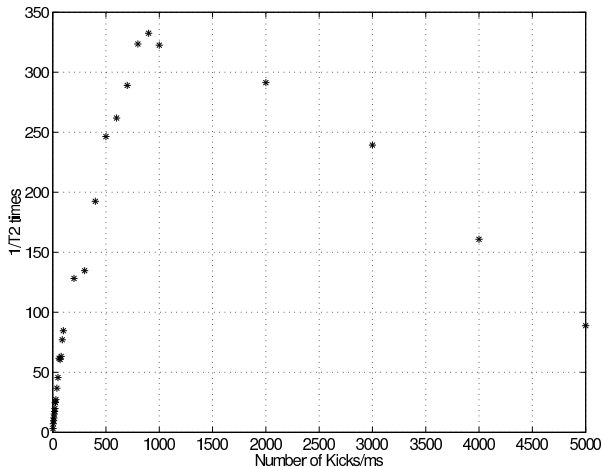}
\caption{Simulation showing the dependence of the decay rate on the kick rate, and the onset of the decoupling limit. Beyond 900 kicks/ms the decay rate decreases.}
\label{fig:kicks}
\end{center}
\end{figure}

A type-II quantum computer is a hybrid classical/quantum device that can potentially 
solve a class of classical computational problems \cite{c58,c59}.  It is essentially an array of 
small quantum information processors sharing information through classical channels.  
Such a lattice of parallel quantum information processors can be mapped onto a liquid 
state NMR system by mapping the lattice sites of the quantum computer onto spatial 
positions in the nuclear spin ensembles. The implementation of a type-II quantum 
computer using NMR techniques has been demonstrated in solving the diffusion equation 
\cite{c60}.  The experimental results show good agreement with both the analytical solutions 
and numerical NMR simulations. The spatial separation of the different lattice sites in the 
ensemble allows one to address all the lattice sites simultaneously using frequency 
selective RF pulses and a magnetic field gradient.  This yields a significant savings in 
time compared to schemes where the sites have to be addressed individually.  

\section{Contributions to other QC systems}

NMR QIP studies have contributed significantly to enabling quantum
computation with other physical systems.  Fundamentally, this has been
because of the exquisite level of control achievable in NMR, which
remains unrivaled.  Several of these contributions are briefly
summarized below; a complete discussion is available in the literature
\cite{Vandersypen04a}.

\subsubsection{Composite pulses: trapped Ions}

The use of {\em composite pulses} has been an important contribution of NMR to QIP.
A single, imperfect pulse is replaced
by a sequence of pulses which accomplishes the same operation with less error.
Historically, in the art of NMR, such sequences were first invented to
compensate for apparatus imperfections, such as frequency offsets and
pulse amplitude miscalibrations.  For example, the machine may perform
rotations 
\be
	\tilde{R}_{\hat n}(\theta) 
	= \exp \lbL{ - i (1+\epsilon) \hat{n} \cdot \vec{I}}\rb
\,,
\label{eq:roterr}
\ee
where $\epsilon$ is an unknown, systematic pulse amplitude error.
Ideally, $\epsilon$ is zero, but in practice, it may vary
geometrically across a sample, or slowly, with time.  Using average
gate fidelity as an error metric, this pulse can be shown to have
error which grows quadratically with $\epsilon$.  In comparison,
consider the sequence
\be
	BB1_{\theta} = 
		   \tilde{R}_{\phi}(\pi)
		   \tilde{R}_{3\phi}(3\pi)
		   \tilde{R}_{\phi}(\pi)
               \tilde{R}_x(\theta) 
\,,
\label{eq:bb1def}
\ee
where $\tilde{R}_{\phi}(\cdot)$ denotes a rotation about the axis
$[\cos\phi,\sin\phi,0]$, and the choice $\phi =
\cos^{-1}(-\theta/4\pi)$ is made.  This sequence, introduced by
Wimperis \cite{Wimperis94a}, gives average gate fidelity error $\sim 21
\pi^6 \epsilon^6/16384$, which is much better than the $O(\epsilon^2)$
for the single pulse, even for relatively large values of $\epsilon$.
Generalizations and extensions of this technique can help correct not
just systematic single qubit gate errors, but also coupled gate
errors \cite{c14,Jones03b}.

NMR composite pulses have also recently been successfully employed in
quantum computation with trapped ions.  In an experiment with a single
trapped $^{40}$Ca ion, a sequence of $O(10)$ laser pulses was
performed using a variety of phases, to implement a proper swap
operation between the internal atomic state and the motional state of
the ion, and a controlled-phase gate.  These steps allowed the full
two-qubit Deutsch-Jozsa algorithm to be implemented \cite{Gulde03a}.
Composite pulses have also been used in superconducting qubits demonstrating
robustness against detuning in a quantronium circuit \cite{Collin04}.

\subsubsection{Shaped pulses: superconducting qubits}

NMR also widely employs shaped pulses to achieve desired control
excitations of the spins.  Typically, this shaping is performed in the
amplitude and phase domain.  One goal of this method, for example, is
to achieve narrow excitation bandwidths.  To first order, the excited
bandwidth is the Fourier transform of the temporal width of the pulse.  
However, because of the nonlinear {\em Bloch} response of the spins to
the RF excitation, the first order approximation rapidly breaks down
for more than small tip angles \cite{Pauly91a}.  
 Thus, in order to achieve sharp
  excitation bandwidths for different tip angles, or uniform
  excitation of the spins over a certain frequency range, a panoply
  of pulse shapes, such as gaussians, hermite-gaussians \cite{Warren84a}, and
  fancifully named ones, including BURP and REBURP \cite{Green91a}  have been
  designed.

These NMR techniques are applicable to precise control of quantum
systems other than NMR.  Numerical optimization can be used to sculpt
pulse shapes to provide desired unitary
transformations \cite{c25},  and shaped pulses may be useful
for controlling Josephson junction phase qubits \cite{Steffen03c}.

\section{Transition to solid state NMR}
While the liquid state studies have allowed us to explore open system dynamics and to 
develop means and metrics for obtaining control in small quantum systems, these studies 
have generally been limited to less than 10 qubits. Though the decoherence times are 
long (on the order of seconds), the strength of the spin-spin coupling (used to implement 
2-qubit gates) is small (about 100 Hz), limiting the number of operations that can be 
performed.  In addition, at room temperature the density matrix characterizing the spin 
system is highly mixed and it is necessary to use pseudo-pure states \cite{Cory97a,Gershenfeld97a}.  As the room 
temperature polarization of the sample is very small ($<$ 1 part in $10^5$) , the exponential 
loss in signal as the size of the spin system grows limits the number of qubits that can be 
observed.  Another limitation to scaling liquid state NMR techniques is the use of 
chemistry for frequency-dependent addressing.  As the number of qubits increases, the 
number of transitions that need to be individually addressed grows as well.  These 
transitions all lie within a fixed (chemistry dependent) bandwidth, making it 
progressively harder to address a single transition without disturbing any other.  While 
techniques such as algorithmic cooling \cite{Schulman98a,c64} and cellular automata \cite{c65} 
schemes have been proposed to overcome some of these limitations, their experimental feasibility has 
not been demonstrated to date. 

Solid state NMR approaches allow us to obtain control over a much larger Hilbert 
space, and hold great promise for the study of many body dynamics and quantum 
simulations.  The most important spin-spin interaction is the through-space dipolar 
coupling, which is on the order of tens of kilohertz in typical dielectric crystals, so that it 
should be possible to implement a large number of operations (perhaps $10^4$) before the 
spins decohere.  Moreover, in the solid state, the spins can be highly polarized by 
techniques such as polarization transfer from electronic spins \cite{c11}.  The increased 
polarization allows an exploration of systems with a larger number of qubits, and also 
allows preparation of the system  close to a pure state.  While traditional solid state NMR 
techniques rely on chemistry for addressing,  it is possible to introduce spatial addressing 
of the spins using extremely strong magnetic field gradients that produce distinguishable 
Larmor precession frequencies on the atomic scale, or via an auxiliary quantum system 
such as an electron spin, a quantum dot or even a superconducting qubit that is coupled to 
the nuclear spin system.  For instance, entanglement between an electron and nuclear spin 
in an ensemble has recently been demonstrated \cite{c66}.

A variety of architectures have been proposed for solid state NMR quantum 
computing, a few of which are enumerated below :
\begin{enumerate}
\item Cory et al.\ proposed an ensemble solid state NMR quantum computer, using a 
large number of $n$-qubit quantum processor molecules embedded in a lattice \cite{c5}.  
The processors are sufficiently far apart that they only interact very weakly with 
each other.  The bulk lattice is a deuterated version of the QIP molecule, with no 
other spins species present.  Paramagnetic impurities in the lattice are used to 
dynamically polarize the deuterium spins, and this polarization can then be 
transferred to the QIP molecules using polarization transfer techniques.  The 
addressing is based on the chemistry of the processor molecule.
\item Yamamoto and coworkers have proposed another ensemble NMR processor, 
using an isotopically engineered silicon substrate containing $^{29}$Si spin chains in 
an $^{28}$Si or $^{30}$Si lattice \cite{c8,c9}.  The $^{29}$Si has spin 1/2 while $^{28}$Si 
and $^{30}$Si have spin 0.
A microfabricated (dysprosium) ferromagnet is used to produce extremely 
strong magnetic field gradients to create a variation of the nuclear spin Larmor 
frequencies at the atomic scale.  Detection is performed using magnetic 
resonance force-microscopy.
\item The use of an N-V defect center in diamond coupled to a cluster of $^{13}$C 
spins as a quantum processor has been proposed by Wrachtrup and coworkers 
\cite{c6}.  The hypefine coupling between the electron spin in the defect and the 
carbon nuclei allows these carbon nuclei to be addressed individually.  Using a 
combination of optically detected electron nuclear double resonance and single 
molecule spectroscopy techniques, they suggest that it should be possible to 
both prepare pure spin states of the system as well as directly detect the result of 
a computation by performing a single spin measurement.
\item Suter and Lim propose an architecture based on endohedral fullerenes, 
encapsulating either a phosphorus or nitrogen atom, positioned on a silicon 
surface \cite{c10}.  The C$_{60}$ cages form atom traps, and the decoherence time of the 
phosphorus or nitrogen nuclear spin is consequently very long.  Switched 
magnetic field gradients that can produce observable electron Larmor frequency 
shifts on nanometer length scales, in combination with frequency selective RF 
pulses are used to address the spins.  Each site represents two physical qubits, 
the electron spin of the fullerene and the nuclear spin of the trapped atom, and 
are used to create a single logical qubit.  Two qubit gates between different 
fullerene molecules are mediated by the electron-electron dipolar coupling.
\end{enumerate}

While simple gates have been demonstrated in solid state NMR \cite{c67}, the fidelity of these 
gates is low, and significant experimental challenges remain.  The decoherence 
mechanism in the solid state is primarily due to the indistinguishability of the chemically 
equivalent nuclear spins.  The Hamiltonian of a homonuclear spin system is \cite{c11}
\begin{equation}
{\cal H} = \omega\sum_i I_z^i + \sum_{i<j} d_{ij} \left(3I_z^iI_z^j- I^i \cdot I^j\right)
\end{equation}
where the first term corresponds to the Zeeman energy and second term is the truncated 
dipolar Hamiltonian in strong magnetic fields.  While the total dipolar Hamiltonian 
commutes with the total Zeeman Hamiltonian, the Zeeman and dipolar terms do not 
commute on a term by term basis.  The phase memory of the spins is scrambled as they 
undergo energy-conserving spin flips with other spins in the system.  Control of the 
dipolar interaction is therefore an essential element of any NMR solid state proposal.
An important example of the precision of control is the ability to effectively suppress 
all internal Hamiltonians and preserve quantum information.  Figure \ref{fig:fid}(a) shows the 
measured free induction decays for a crystal of calcium fluoride at two different crystal 
orientations, while Figure \ref{fig:fid}(b) shows the signal while refocusing the dipolar interaction. 
It is therefore possible to experimentally extend the coherence times of the $^{19}$F spins in a 
single crystal of calcium fluoride from 100 $\mu$s (no modulation), to 2 ms using standard 
NMR techniques, and finally to 500 ms using recently developed methods \cite{c68}.  This 
represents an increase by approximately 4 orders of magnitude.  In the limit of perfect coherent 
control of the dipolar couplings, it should be possible to significantly further extend the 
coherence time of the spins.  In addition to improving coherent control, such studies 
provide insight into the next important contribution to decoherence.

\begin{figure}
\begin{center}
\includegraphics[scale=0.35]{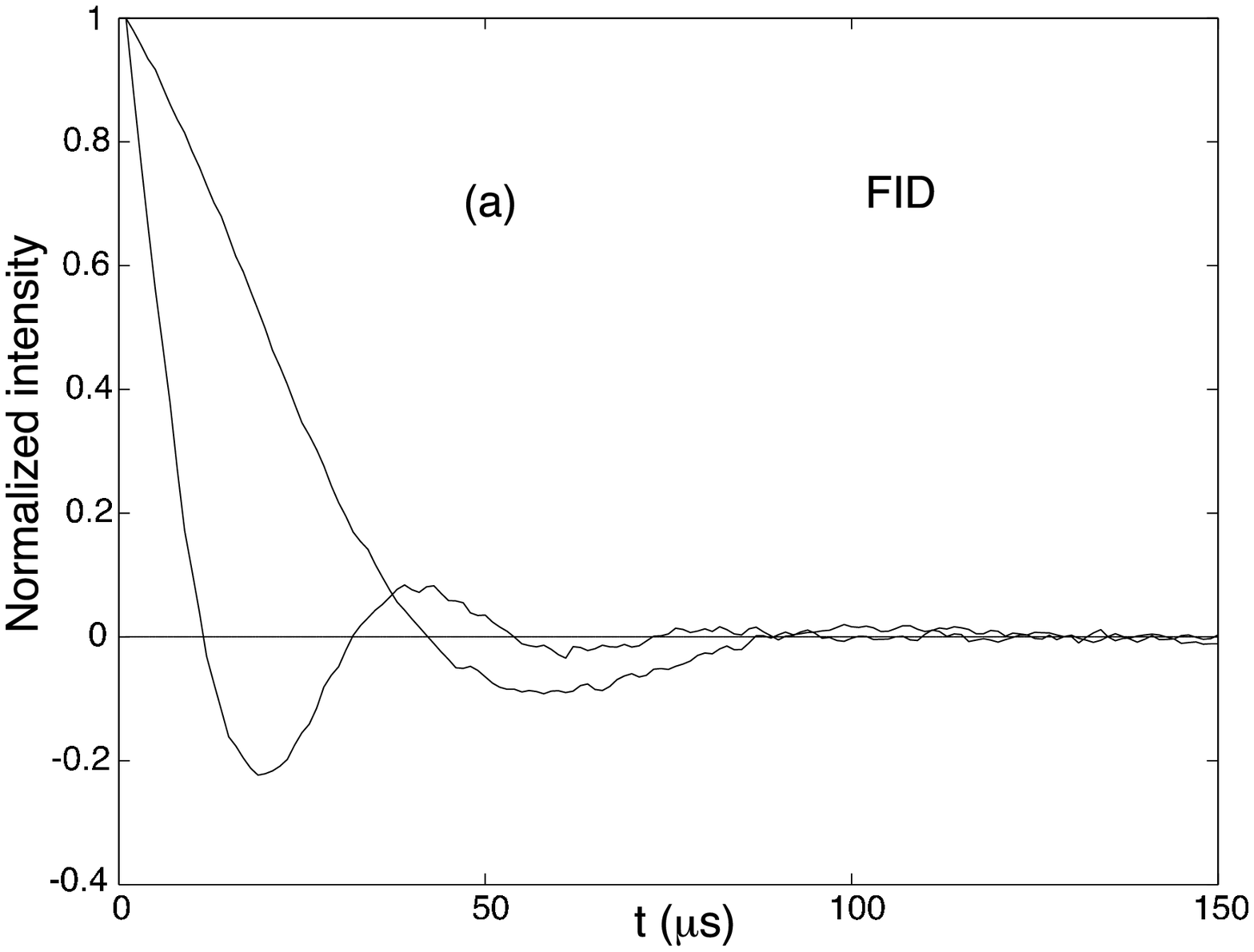}
\includegraphics[scale=0.35]{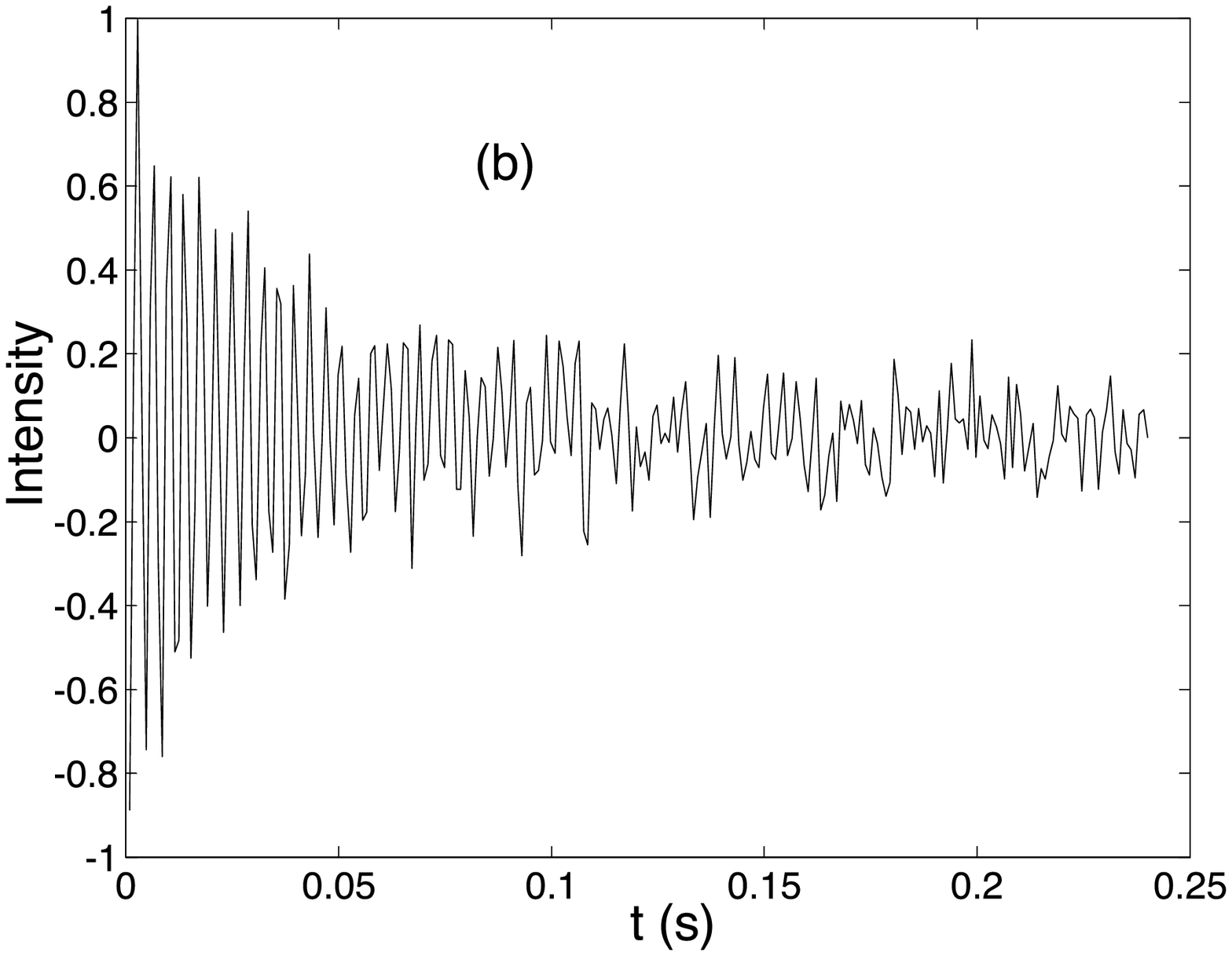}
\caption{(a) Free induction decay of a single crystal of calcium fluoride, at two different crystal 
orientations (b) Decay of the $^{19}$F signal while the dipolar coupling between the nuclear spins is 
decoupled. (The sequence was modified to show an oscillation, so the decoupling is not optimal 
here.)}
\label{fig:fid}
\end{center}
\end{figure}

The decay of the observed FID in Figure \ref{fig:fid}(a) is due to the mutual dipolar 
couplings of the spins.  These couplings produce correlated many spin states that are not directly 
observable using standard NMR techniques.  However, using multiple quantum encoding 
techniques \cite{c69}, it is possible to directly measure the growth of the spin system under the 
dipolar coupling. The truncated dipolar Hamiltonian shown above is a zero quantum 
Hamiltonian when examined in the basis of the quantizing Zeeman Hamiltonian as 
expected.  However, we can requantize the system in another basis (such as the $x$-basis 
for example) via a similarity transformation, and explore the growth of multiple quantum 
coherences in this new basis.  The dipolar Hamiltonian in the $x$-basis is 
\begin{equation}
{\cal H} = -\frac{1}{2}\sum_{i<j} d_{ij}\left\{2I_x^i I_x^j - \frac{1}{2}\left(I_{x+}^{i} I_{x-}^{j}+ I_{x-}^{i} I_{x+}^{j}\right)\right\} - \frac{3}{4} \sum_{i<j} d_{ij}\left(I_{x+}^{i} I_{x+}^{j}+ I_{x-}^{i} I_{x-}^{j}\right)
\end{equation}
and is thus seen to contain both zero and double quantum terms.  It is possible to directly 
observe the growth of these $x$-basis coherences. Figure \ref{fig:mqfid} shows the results of this 
experiment, illustrating the growth in the number of the correlated spins from 1 to about 
20 in the first 150 $\mu$s following the application of a $\pi$/2 pulse.

\begin{figure}
\begin{center}
\includegraphics[scale=0.7]{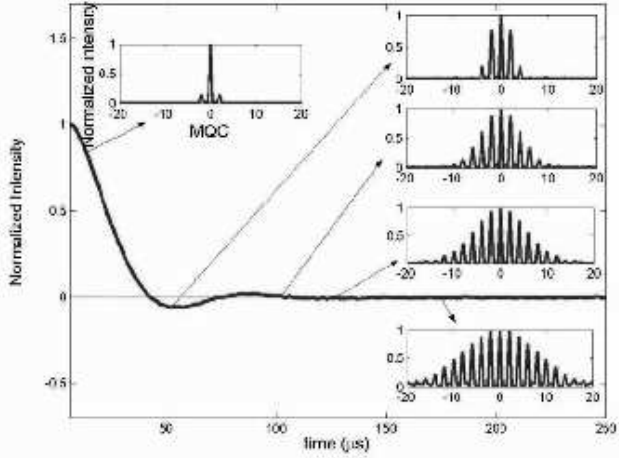}
\caption{Multiple quantum encoding, combined with evolution reversal sequences allow us to 
follow the growth of the correlated spin states during the course of a free induction decay.  It is 
seen that even when the macroscopic signal appears to have decayed away, the spin system 
remains highly coherent, and states involving up to 20 correlated spins are observed.}
\label{fig:mqfid}
\end{center}
\end{figure}

In an early demonstration of the capability to explore many body dynamics in spin 
systems, solid state NMR techniques have been used to directly measure the rate of spin 
diffusion of Zeeman and dipolar energy in a single crystal of calcium fluoride \cite{c70,c71}.  
As seen from the Hamiltonian above, these are both constants of the motion and are 
independently conserved.  Spin diffusion is a coherent process caused by the mutual spin 
flips induced by the dipolar coupling between spins, that appears diffusive in the long-time, 
long-wavelength limit \cite{c72}.  It is estimated that up to $10^{18}$ spins are involved at the 
long timescales explored in these experiments.  It was found that while the measured 
diffusion coefficients for Zeeman order were in good agreement with theoretical 
predictions, the diffusion of dipolar order was observed to be significantly faster than 
previously predicted.   While these experiments were performed in highly mixed thermal 
states, future experiments planned at low temperatures, and high polarizations should 
enable a more complete exploration of the large Hilbert space dynamics.

\section{Conclusions}
NMR implementations of QIP have thus yielded a wealth of information by providing experimental realizations of a number of proposed schemes.  This in turn has guided our understanding of the relevant issues involved in scaling these testbed systems up in size.  The methodologies developed are relevant across most of the physical platforms that have been proposed for QIP, and manifestations of this "cross-fertilization" are beginning to appear in the literature.  

Solid state NMR holds great promise for scalable QIP architectures.  The efforts currently underway to characterize and control large spin systems are essential to determining how the methodologies of control and system decoherence scale as a function of the system Hilbert space size.   

\section{Acknowledgements}
This work was supported by funds from ARDA/ARO (DAAD19-01-1-0519), DARPA (MDA972-01-1-0003), the NSF (EEC 0085557) and the Air Force Office of Sponsored Research.

\end{document}